\documentstyle[osa,manuscript]{revtex}

\begin{document}
\title{{\small {\it Foundations of Physics Letters {\bf 11}(2), 165-178, 1998.
\bigskip \bigskip \bigskip }} \\
{\normalsize {\bf INCOMPLETE DELTA FUNCTIONS \bigskip}} }
\author{{\normalsize {\bf Alexander Gersten \bigskip }}}
\address{{\normalsize {\it Department of Physics and Unit of Bioengineering}} \\
{\normalsize {\it Ben-Gurion University of the Negev \thanks{}1]{Permanent
address.} }} \\
{\normalsize {\it Beer-Sheva 84105, Israel}} \\
{\normalsize and} \\
{\normalsize {\it Department of Physics}} \\
{\normalsize {\it City College of the City University of New York}} \\
{\normalsize {\it New York, NY 10031 \bigskip \bigskip}}}
\date{{\normalsize Received 3 January 1997 \bigskip \bigskip}}
\maketitle

{\normalsize By applying projection operators to state vectors of
coordinates we obtain subspaces in which these states are no longer
normalized according to Dirac's delta function but normalized according to
what we call "incomplete delta functions". We show that this class of
functions satisfy identities similar to those satisfied by the Dirac delta
function. The incomplete delta functions may be employed advantageously in
projected subspaces and in the link between functions defined on the whole
space and the projected subspace. We apply a similar procedure to finite
dimensional vector spaces for which we define incomplete Kronecker deltas.
Dispersion relations for the momenta are obtained and "sums over poles" are
defined and obtained with the aid of differences of incomplete delta
functions. \bigskip \bigskip }


\section{INTRODUCTION}

Projection operators, introduced to quantum mechanics by Dirac [1], are
extremely useful tool in nonrelativistic ${[2,3]}$ and relativistic quantum
mechanics ${[4]}$.

In this work we found that by projecting to coordinate or momenta subspaces,
instead of the Dirac's delta function normalization of state vectors, we
obtain normalizations according to what we call "incomplete delta
functions." These functions can converge, by a limiting procedure, to the
Dirac delta function. Moreover this class of functions satisfy identities
similar to those satisfied by the Dirac delta function. This way the
identities remain the same during the convergence procedure. The identities
so obtained can be used to check hypotheses on functions that their support
is restricted only to the subspace.

As is well known, the delta function is a distribution ${[5]}$ or
(equivalently) a generalized function ${[6]}$, hence the incomplete delta
functions will also be in general distributions. In the following sections
by functions we will also mean distributions. I.M.~Gelfand ${[7]}$ justified
the Dirac bra-ket formalism in the rigged Hilbert space. Also Arye Friedman $%
{[8]}$ has shown that the Dirac bra-ket formalism is adequate, even if the
results are infinite, if nonstandard analysis and nonstandard logic are
used. Therefore we will exclude from our considerations problems of possible
infinities as they should not violate the Dirac formalism. Principal value
integrations will be used throughout the paper.

\section{DEFINITIONS AND GENERAL PROPERTIES}

Let us consider two complete continuous vector bases $|x\rangle $ and $%
|p\rangle $, in the same space, where x are coordinates and p momenta
respectively, normalized according to 
$$
\langle x|x^{\prime}\rangle = \delta ( x - x^{\prime}), \: \: \: \: \langle
p|p^{\prime}\rangle = \delta ( p - p^{\prime}), \: \: \: \:  - \infty \le p
\le \infty , \: \: \: \: - \infty \le x \le \infty , \eqno(2.1) 
$$
and linked together by the wavefunction ($\hbar =1$ will be assumed
throughout the paper) 
$$
\langle x|p\rangle = \frac{1}{\sqrt{2\pi}} \exp (ixp), \eqno(2.2) 
$$
where $\delta (x)$ is the delta function of Dirac. The completeness relation
is given by 
$$
I=\int_{-\infty}^{\infty} |x\rangle \langle x|dx = \int_{-\infty}^{\infty} 
|p\rangle \langle p|dp, \eqno(2.3) 
$$
where I is the identity operator ${[1]}$ . Let us consider the following set
of functions 
\[
\delta (x-x^{\prime},a)=\int_{-a}^{a} \langle x|p\rangle \langle
p|x^{\prime}\rangle dp =\frac{1}{2\pi}\int_{-a}^{a} e^{ip(x-x^{\prime})}dp 
\]
$$
=\frac{a}{\pi}\frac{\sin (a(x-x^{\prime}))}{a(x-x^{\prime})} , \: \: \: \:
\qquad a>0 \eqno(2.4) 
$$
\vspace{.15in} from which the Dirac delta function $\delta (x)$ can be
obtained in the limiting process 
$$
\delta (x) = \lim_{a \to \infty}\delta (x,a) . \eqno(2.5) 
$$

The Dirac delta function has the following properties: 
$$
\delta (x) = \left\{ 
\begin{array}{cc}
0, & \: \: \: \: for \; x\neq 0 \\ 
\infty , & \: \: \: \: for \; x=0.
\end{array}
\right. ,\quad \int_{-\infty}^{\infty} \delta (x) = 1, \eqno(2.6a) 
$$
$$
\int_{-\infty}^{\infty} \delta (x_1 - x)\delta (x-x_0)dx = \delta (x_1 -
x_0), \eqno(2.6b) 
$$
$$
\int_{-\infty}^{\infty} f(x)\delta (x-x_0)dx=f(x_0) \eqno(2.6c) 
$$

We will generalize Eqs. (2.6b) and (2.6c) to a larger class of functions
which we will call "incomplete delta functions."  Let P be a projection
operator, such that by changing its parameters, in the limiting process, it
converges to the identity operator. An example is Eq. (2.4) , in which the
following projection operator was used: 
$$
P(a)=\int_{-a}^a |p\rangle \langle p|dp,\; \: \: \: \:
\lim_{a\to\infty}P(a)=I. \eqno(2.7) 
$$

The projection operator is equal to its square: 
$$
P=P^2 \eqno(2.8a) 
$$
and also equal to its adjoint: 
$$
P=P^{\dagger}, \eqno(2.8b) 
$$
a property which can be employed to derive general properties of the
incomplete delta functions. The proof of Eq. (2.8b) is simple: From Eq.
(2.8a) we have 
$$
P^{\dagger}=(P^{\dagger})^2, \eqno(2.8c) 
$$
While Eqs. (2.8a) and (2.8c) give $P(P^{\dagger})^2=P^2P^{\dagger}$,  from
which (2.8b) results.

Let us define the incomplete delta function, linked to the projection
operator P, as 
$$
\delta (x,x^{\prime},P)=(\langle x|P^{\dagger})(P|x^{\prime}\rangle
)=\langle x|P|x^{\prime}\rangle , \eqno(2.9) 
$$
$$
\delta (x,x^{\prime},P\to I)=\delta (x-x^{\prime}). \eqno(2.9a) 
$$
Above, the states denoted by x, are projected on a subspace by the
projection operator P. We can prove that equations similar to Eqs.
(2.6b-2.6c) are satisfied for the incomplete delta functions. Inserting the
identity operator of Eq. (2.3) into Eq. (2.9) we obtain: 
$$
\begin{array}{ll}
\delta (x,x^{\prime},P) & = {\displaystyle (\langle
x|P)I(P|x^{\prime}\rangle )=\int_{-\infty}^{\infty} \langle
x|P|x^{\prime\prime} \rangle \langle x^{\prime\prime} |P|x^{\prime}\rangle
dx^{\prime\prime} } \\[.1in] 
& {\displaystyle =\int_{-\infty}^{\infty} \delta (x,x^{\prime\prime}
,P)\delta (x^{\prime\prime} ,x^{\prime},P)dx^{\prime\prime} .}
\end{array}
\eqno(2.10) 
$$
Let f(x) be a wavefunction with a support in the projected subspace; than 
$$
\begin{array}{ll}
f(x) & ={\displaystyle \langle x|f\rangle =\langle x|P|f\rangle =\langle
x|PIP|f\rangle } \\[.1in] 
& ={\displaystyle \int_{-\infty}^{\infty} \langle x|PP|x^{\prime}\rangle
\langle x^{\prime}|P|f\rangle dx^{\prime}} \\[.2in] 
& ={\displaystyle \int_{-\infty}^{\infty} \delta
(x,x^{\prime}P)f(x^{\prime})dx^{\prime}.}
\end{array}
\eqno(2.11) 
$$
For all wavefunctions, the following identity holds: 
$$
\langle x|P|f\rangle =\int_{-\infty}^{\infty} \delta (x,x^{\prime},P)\langle
x^{\prime}|P|f\rangle dx^{\prime}. \eqno(2.12) 
$$
The identities (2.11) may have applications in testing hypotheses that a
given function has a support in a restricted subspace. Equations. (2.10) and
(2.12) tend to Eqs. (2.6b) and (2.6c) when the projection operator P tends
to the identity operator I.

\section{EXAMPLES}

Let us elaborate in some detail the properties of some of the incomplete
delta functions.

{\raggedright {\bf Example 1.}} The incomplete delta function linked to the
projection operator $P(a)$ of eq. (2.7) was evaluated in Eq. (2.4). In
section 2 we used 
$$
P(a)=\int_{-a}^a |p\rangle \langle p|dp;\; \langle x|p\rangle  =\frac{1}{%
\sqrt{2\pi}} e^{ixp} , \eqno(3.1) 
$$
and found 
$$
\delta (x,x^{\prime},P(a))\equiv\delta (x-x^{\prime},a)=\frac{\sin
(a(x-x^{\prime}))}{\pi (x-x^{\prime})} ,  \: \: \: \: a>0. \eqno(3.2) 
$$
\vspace{.15in} Equations (2.10) and (2.12) [similar to Eqs. (2.6b-2.6c)]
take the form: 
$$
\frac{\sin (a(x-x^{\prime}))}{\pi (x-x^{\prime})} =\int_{-\infty}^{\infty} 
\frac{\sin (a(x-x^{\prime\prime} ))}{\pi (x-x^{\prime\prime} )}\frac{\sin
(a(x^{\prime\prime} -x^{\prime}))}{\pi (x^{\prime\prime} -x^{\prime})}
dx^{\prime\prime} , \eqno(3.3) 
$$
$$
\langle x|P(a)|f\rangle =\int_{-\infty}^{\infty}\frac{\sin (a(x-x^{\prime}))%
}{\pi (x-x^{\prime})} \langle x^{\prime}|P(a)|f\rangle dx^{\prime}. \eqno%
(3.4) 
$$
\vspace{.15in}

In order to become more familiar with Eq. (3.4), let us work out a simple
example and consider $f(x)=\exp (ibx)$ . We obtain: 
\[
\displaystyle
\begin{array}{ll}
\langle x|P(a)|f\rangle & ={\displaystyle \int_{-a}^a \langle x|p\rangle
\langle p|f\rangle dp} \\[.1in] 
& ={\displaystyle \int_{-\infty}^{\infty} dx^{\prime}\int_{-a}^a \langle
x|p\rangle \langle p|x^{\prime}\rangle \langle x^{\prime}|f\rangle dp} \\%
[.1in] 
& ={\displaystyle \frac{1}{2\pi}\int_{-a}^a dp\int_{-\infty}^{\infty}
dx^{\prime}\exp \left[ ip(x-x^{\prime})+ibx^{\prime}\right]} \\[.2in] 
& ={\displaystyle \int_{-a}^a dp\exp (ipx)\delta (p-b)} \\[.2in] 
& =\left\{ 
\begin{array}{cc}
\exp (ibx), & \: \: \: \: |b|\le a, \\[.1in] 
0, & \: \: \: \: |b|>a,
\end{array}
\right.
\end{array}
\]
Substituting this result in Eq. (3.4) we obtain 
$$
\int_{-\infty}^{\infty}\frac{\sin (a(x-x^{\prime}))}{\pi (x-x^{\prime})}
\exp (ibx^{\prime})dx^{\prime}=\exp (ibx),\: \: \: \: |b|\le a. \eqno(3.5) 
$$
We shall prove the following relation satisfied by the incomplete delta
function of Eq. (3.1): 
$$
\delta (bx,a)=\frac{1}{|b|}\delta (x,a|b|), \eqno(3.6) 
$$
where b is a real number. For $b>0$ we find
$$
\delta (bx,a)=\frac{1}{2\pi}\int_{-a}^a e^{ixby} dy = \frac{1}{2b\pi}
\int_{-ab}^{ab} e^{ixz} dz = \frac{1}{b} \delta (x,ab). \eqno(3.7) 
$$
For $b<0,$ we obtain 
\[
\begin{array}{ll}
\delta (bx,a) & ={\displaystyle \frac{1}{2\pi}\int_{-a}^a e^{ixby} dy =\frac{%
1}{2b\pi}\int_{-ab}^{ab} e^{ixz} dz } \\[.1in] 
& ={\displaystyle \frac{1}{2|b|\pi}\int_{-a|b|}^{a|b|} e^{ixz} dz } \\[.2in] 
& ={\displaystyle \frac{1}{|b|} \delta (x,a|b|),}
\end{array}
\]
which, together with Eq. (3.7), proves Eq. (3.6). From Eq. (3.6), we can
get, in the limiting process, the wellknown result 
$$
\delta (bx) = \frac{1}{|b|}\delta(x). \eqno(3.8) 
$$

{\raggedright {\bf Example 2.}} The incomplete delta function obtained with
the operator P(a) given by Eq. (2.7) and acting between the p basis vectors
has the following form: 
$$
\begin{array}{ll}
\delta (p,p^{\prime},P(a)) & ={\displaystyle \langle
p|P(a)|p^{\prime}\rangle =\int_{-a}^a \langle p|p^{\prime\prime} \rangle
\langle p^{\prime\prime} |p^{\prime}\rangle dp^{\prime\prime} } \\[.1in] 
& = \left\{ 
\begin{array}{cc}
\delta (p-p^{\prime}), & \: \: \: \: if\; |p|\le a \; and \; |p^{\prime}|\le
a, \\[.1in] 
0, & \: \: \: \: if \; |p|>a \; or \; |p^{\prime}|>a,
\end{array}
\right.
\end{array}
\eqno(3.9) 
$$

{\raggedright {\bf Example 3.}} The incomplete delta functions in the x
space, linked with projections into a half axis of the p space: 
$$
P_+=\int_{0+}^{\infty} |p\rangle \langle p|dp, \: \: \: \:
P_-=\int_{-\infty}^{0-}, \: \: \: \: I=P_++P_-, \: \: \: \: P_+P_-=0, \eqno%
(3.10) 
$$
have the following form: 
$$
\begin{array}{ll}
\delta (x,x^{\prime},P_+) & ={\displaystyle \langle x|P_+|x^{\prime}\rangle
=\int_{0+}^{\infty} \langle x|p\rangle \langle p|x^{\prime}\rangle dp } \\%
[.1in] 
& ={\displaystyle \frac{1}{2\pi}\int_{0+}^{\infty} \exp (ip(x-x^{\prime}))dp}
\\[.1in] 
& ={\displaystyle \lim_{\epsilon\to 0}\frac{1}{2\pi}\int_{0+}^{\infty}\exp
(ip(x-x^{\prime})-\epsilon p)dp } \\[.1in] 
& ={\displaystyle \lim_{\epsilon\to 0}\frac{-1}{2\pi i}\frac{1}{%
x-x^{\prime}+i\epsilon}} ={\displaystyle \lim_{\epsilon\to 0}\frac{-1}{2\pi i%
}\frac{x-x^{\prime}-i\epsilon}{(x-x^{\prime})^2+ \epsilon^2}} \\[.15in] 
& = {\displaystyle\frac{-1}{2\pi i}\frac{1}{x-x^{\prime}} +\lim_{\epsilon\to
0}\frac{1}{2\pi}\frac{\epsilon}{(x-x^{\prime})^2+\epsilon^2}} \\[.15in] 
& ={\displaystyle \frac{-1}{2\pi i}\frac{1}{x-x^{\prime}}+\frac{1}{2}\delta
(x-x^{\prime}), }
\end{array}
\eqno(3.11) 
$$
$$
\begin{array}{ll}
\delta (x,x^{\prime},P_-) & ={\displaystyle \langle x|P_-|x^{\prime}\rangle
=\langle x|I-P_+|x^{\prime}\rangle =\delta (x-x^{\prime})-\langle
x|P_+|x^{\prime}\rangle } \\[.1in] 
& ={\displaystyle \frac{1}{2\pi i}\frac{1}{x-x^{\prime}}+\frac{1}{2}\delta
(x-x^{\prime}),}
\end{array}
\eqno(3.12) 
$$
$$
\delta (x,x^{\prime},P_+)+\delta (x,x^{\prime},P_-)=\langle
x|P_++P_-|x^{\prime}\rangle =\langle  x|x^{\prime}\rangle =\delta
(x-x^{\prime}). \eqno(3.13) 
$$
Eq. (2.10), takes for this example, the form 
$$
\begin{array}{ll}
{\displaystyle \frac{-1}{2\pi i}\frac{1}{x-x^{\prime}} } & +\frac{1}{2}%
\delta (x-x^{\prime}) \\[.1in] 
& ={\displaystyle \int_{-\infty}^{\infty} \left[ \frac{-1}{2\pi i} \frac{1}{%
x-x^{\prime\prime} }+\frac{1}{2}\delta (x-x^{\prime\prime} ) \right] \left[ 
\frac{-1}{2\pi i}\frac{1}{x^{\prime\prime} -x^{\prime}}+\frac{1}{2}\delta
(x^{\prime\prime} -x^{\prime}) \right] dx^{\prime\prime} }
\end{array}
\eqno(3.14) 
$$
from which we infer that 
$$
\int_{-\infty}^{\infty}\frac{dx^{\prime\prime} }{(x-x^{\prime\prime}
)(x^{\prime\prime} -x^{\prime})}=-\pi^2\delta (x-x^{\prime}). \eqno(3.15) 
$$
\vspace{.15in} A consequence of Eq. (3.15) is a general relation for a
function $f(x)$: 
$$
f(x)=\int_{-\infty}^{\infty}\delta (x-x^{\prime})f(x^{\prime})dx^{\prime}=%
\frac{-1}{\pi^2}\int_{-\infty}^{\infty}\int_{-\infty}^{\infty} \frac{%
f(x^{\prime})dx^{\prime}dx^{\prime\prime} }{(x-x^{\prime\prime}
)(x^{\prime\prime} -x^{\prime})}. \eqno(3.15a) 
$$
Eq. (2.12), for example 3, takes the form 
$$
\langle x|P_+|f\rangle = \int_{-\infty}^{\infty} \left[ \frac{-1}{2\pi i}%
\frac{1}{x-x^{\prime}} +\frac{1}{2}\delta (x-x^{\prime}) \right]  \langle
x^{\prime}|P_+|f\rangle dx^{\prime}, \eqno(3.16) 
$$
or 
$$
\langle x|P_+|f\rangle =\frac{1}{\pi i}\int_{-\infty}^{\infty}\frac{\langle
x^{\prime}|P_+|F\rangle }{x^{\prime}-x} dx^{\prime}, \eqno(3.16a) 
$$
\vspace{.15in} i.e., the real and imaginary parts of $\langle x|P_+|f\rangle 
$ are connected  via a Hilbert transform ${[10]}$. As an example of an
application of Eq. (3.16), let us derive, in a simplified way, an indirectly
known relation ${[9]}$ for the spectral amplitude of the signal 
$$
f(t)= \left\{ 
\begin{array}{cc}
u(t), & \: \: \: \: for \; t\ge 0 \\ 
0, & \: \: \: \: for \; t<0
\end{array}
\right. \eqno(3.17) 
$$
The above formulas will be used with the replacements 
$$
y\to t, \: \: \: \: x\to -\omega, \eqno(3.18) 
$$
where $t$ is the time and $\omega$ the angular frequency. The transition
matrix between these two bases is 
$$
\langle \omega |t\rangle =\exp (-i\omega t)/\sqrt(2\pi). \eqno(3.19) 
$$
From Eqs. (3.16a), (3.17), and (3.18) we obtain 
$$
\langle \omega |P_+|f\rangle =\langle \omega |f\rangle = \frac{-1}{i\pi} 
\int_{-\infty}^{\infty} \frac{\langle \omega ^{\prime}|f\rangle }{\omega
^{\prime}-\omega} d\omega ^{\prime}, \eqno(3.20) 
$$
i.e., the imaginary and real parts of the spectral amplitude are
interrelated via the Hilbert transform.

{\raggedright {\bf Example 4.}} The incomplete delta function obtained with
the operator $P_+$ given by Eq. (10) and acting between the p basis vectors
has the form 
$$
\begin{array}{ll}
\delta (p,p^{\prime},P_+) & ={\displaystyle \langle p|P_+|p^{\prime}\rangle
= \int_{0+}^{\infty} \langle p|p^{\prime\prime} \rangle \langle
p^{\prime\prime} |p^{\prime}\rangle dp^{\prime\prime} } \\[.1in] 
& = \left\{ 
\begin{array}{cc}
\delta (p-p^{\prime}), & \: \: \: \: if \; p\ge 0 \; and \; p^{\prime}\ge 0
\\ 
0, & \: \: \: \: if \; p<0 \; or \; p^{\prime}<0.
\end{array}
\right.
\end{array}
\eqno(3.21) 
$$

\section{THE INCOMPLETE KRONECKER DELTAS}

In Secs. 2 and 3 we have treated incomplete delta functions in infnite
dimensional spaces; here, in order to visualize better the finite
dimensional case we will treat finite samples of signals with their finite
Fourier transforms. We can represent a sampled signal $u(t_k) \;
(k=1,2,...,K)$ by the vector 
$$
|u\rangle = \left( 
\begin{array}{c}
u(t_1) \\ 
u(t_2) \\ 
\vdots \\ 
u(t_k)
\end{array}
\right) \eqno(4.1) 
$$
in the Kdimensional vector space spanned by the orthonormal base 
$$
|t_1\rangle = \left( 
\begin{array}{c}
1 \\ 
0 \\ 
\vdots \\ 
0
\end{array}
\right) , \: \: \: \: |t_2\rangle = \left( 
\begin{array}{c}
0 \\ 
1 \\ 
\vdots \\ 
0
\end{array}
\right) , \: \: \: \: \cdots \: \: \: \: |t_k\rangle = \left( 
\begin{array}{c}
0 \\ 
0 \\ 
\vdots \\ 
1
\end{array}
\right) , \eqno(4.2) 
$$
Thus 
$$
\langle t_k|u\rangle =u(k), \eqno(4.3) 
$$
where 
\[
\langle t_k|=|t_k\rangle ^{\dagger}= \left( \overbrace{0,\cdots 0}^{k-1 \;
times} ,1,0,\cdots 0 \right) 
\]
is the adjoint (bra) basis. We may have an angular frequency orthonormal
basis if 
$$
|\omega_k\rangle =\frac{1}{\sqrt{K}} \left( 
\begin{array}{c}
exp(i\omega_k t_1) \\ 
exp(i\omega_k t_2) \\ 
\vdots \\ 
exp(i\omega_k t_k)
\end{array}
\right) , \quad \omega_k = 2\pi f/K, \; k=1,2,\cdots K. \eqno(4.4) 
$$
For both bases the identity operator I will be 
$$
I=\sum_{k=1}^K |t_k\rangle \langle t_k| = \sum_{k=1}^K  |\omega_k\rangle
\langle \omega_k| , \eqno(4.5) 
$$
and the components of the spectral amplitude of the signal are 
$$
\langle \omega_k |u\rangle =\sum_{n=1}^K \langle \omega_k  |t_n\rangle
\langle t_n|u\rangle =\frac{1}{\sqrt{K}}\sum_{n=1}^K u(t_n)\exp (i\omega_k
t_n). \eqno(4.6) 
$$
Both orthonormal bases satisfy the relations 
$$
\langle t_k|t_n\rangle =\delta_{kn} , \quad \langle 
\omega_k|\omega_n\rangle =\delta_{kn} , \eqno(4.7) 
$$
where $\delta_{kn}$ is the Kronecker delta 
$$
\delta_{kn} = \left\{ 
\begin{array}{cc}
1, & \: \: \: \: if \; k=n \\ 
0, & \: \: \: \: if \; k\neq n.
\end{array}
\right. \eqno(4.8) 
$$
Let us project to a subspace with a projection operator P: 
$$
P=\sum_{k=K_1}^{K_2} |\omega_k\rangle \langle \omega_k|, \: \: \: \: 1\le
K_1<K_2<K. \eqno(4.9) 
$$
In the projected subspace the incomplete Kronecker delta will be defined, in
analogy with Eq. (2.9), as 
$$
\begin{array}{ll}
\delta_{kn} (P) & ={\displaystyle \langle t_k|P|t_n\rangle =
\sum_{m=K_1}^{K_2} \langle t_k|\omega_m\rangle \langle \omega_m|t_n\rangle }
\\ 
& ={\displaystyle \frac{1}{K}\sum_{m=K_1}^{K_2}\exp [i\omega_m (t_k - t_m)] }
\\[.22in] 
& ={\displaystyle \frac{1}{K}\frac{\exp [i(K_2+1)\omega_1(t_k-t_n)]-\exp
[iK_1\omega_1(t_k-t_n)]}{\exp [i\omega_1(t_k-t_n)]-1}. }
\end{array}
\eqno(4.10) 
$$
\vspace{.15in} In analogy with Eqs. (2.10) and (2.12), the following
relations will be satisfied: 
$$
\delta_{kn}(P)=\sum_{m=1}^K \delta_{km}(P)\delta_{mn}(P), \eqno(4.11) 
$$
$$
\langle t_k|P|u\rangle =\sum_{m=1}^K\delta_{km}(P)\langle t_m|P|u\rangle . %
\eqno(4.12) 
$$
Equations (4.11) and (4.12) can be generalized to arbitrary projections.

\section{DISPERSION RELATIONS AND SUMS OVER POLES}

In sect. 3, Eq. (3.10), we have introduced the projection operators $P_+$
and $P_-$ , and in Eqs. (3.11) and (3.12) the linked incomplete delta
functions. The derived Eq. (3.16a) is peculiar in interrelating the
imaginary and the real parts. Such a situation is often called a dispersion
relation. In this section we will look for further possibilities of
interrelating real and imaginary parts. In a similar way to Eqs. (3.10) we
can define: 
$$
P^+=\int_{0+}^{\infty} |x\rangle \langle x|dx, \: \: \: \:
P^-=\int_{-\infty}^{0-}  |x\rangle \langle x|dx, \: \: \: \: I=P^++P^-, \:
\: \: \: P^+P^-=0, \eqno(5.1) 
$$
and in a way, similar to the results (3.11) and (3.12), obtain: 
$$
\begin{array}{ll}
\delta(p,p^{\prime},P^+) & ={\displaystyle \langle p|P^+|p^{\prime}\rangle =
\int_{0+}^{\infty} \langle p|x\rangle \langle x|p^{\prime}\rangle dx } \\%
[.1in] 
& ={\displaystyle \frac{1}{2\pi} \int_{0+}^{\infty} \exp
(-ix(p-p^{\prime}))dx } \\[.2in] 
& ={\displaystyle \frac{1}{2\pi i} \frac{1}{p-p^{\prime}} + \frac{1}{2}%
\delta (p-p^{\prime}), }
\end{array}
\eqno(5.2) 
$$
$$
\delta(p,p^{\prime},P^-)=\langle p|P^-|p^{\prime}\rangle =\frac{-1}{2\pi i}%
\frac{1}{p-p^{\prime}}  +\frac{1}{2}\delta (p-p^{\prime}) , \eqno(5.3) 
$$
\vspace{.15in} In a way, similar to obtaining the result (3.16a), we can get
the dispersion relation 
$$
\langle p|P^+|f\rangle =\frac{1}{\pi i}\int_{-\infty}^{\infty}\frac{\langle
p^{\prime}|P^+|f\rangle }{p-p^{\prime}} dp^{\prime}. \eqno(5.4) 
$$
Let us note the following identity, based on Eqs. (2.8a) and (5.1): 
$$
I=P^-+P^+=(P^+-P^-)(P^+-P^-). \eqno(5.5) 
$$
From Eqs. (5.1-5.3), we get 
$$
\langle p|(P^+-P^-)|p^{\prime}\rangle =\frac{1}{\pi i}\frac{1}{p-p^{\prime}}%
. \eqno(5.6) 
$$
from which the following identities can be derived: 
$$
\langle p|(P^+-P^-)|f\rangle =\int_{-\infty}^{\infty} \langle 
p|(P^+-P^-)|p^{\prime}\rangle \langle p^{\prime}|f\rangle dp^{\prime}, \eqno%
(5.7) 
$$
$$
\langle p|f\rangle =\int_{-\infty}^{\infty} \langle
p|(P^+-P^-)|p^{\prime}\rangle  \langle p^{\prime}|(P^+-P^-)|f\rangle
dp^{\prime}. \eqno(5.8) 
$$
Inserting Eqs. (5.1) and (5.6) into Eqs. (5.7) and (5.8), we obtain: 
$$
\begin{array}{ll}
\langle p|(P^+ & -P^-)|f\rangle = {\displaystyle \int_{0}^{\infty} \langle
p|x\rangle \langle x|f\rangle dx -\int_{-\infty}^0 \langle p|x\rangle
\langle x|f\rangle dx} \\[.1in] 
& ={\displaystyle \frac{1}{\sqrt{2\pi}} \int_{0}^{\infty} \exp (-ixp)\langle
x|f\rangle dx - \frac{1}{\sqrt{2\pi}} \int_{-\infty}^0 \exp (-ixp)\langle
x|f\rangle dx } \\[.2in] 
& \qquad ={\displaystyle \frac{1}{\pi i} \int_{-\infty}^{\infty} \frac{%
\langle p^{\prime}|f\rangle }{p-p^{\prime}} dp^{\prime},}
\end{array}
\eqno(5.9) 
$$
which states that the difference of incomplete Fourier transforms is equal
to sum over poles whose residues are the values of the Fourier transformed
wavefunction. On the other hand, from Eqs. (5.6) and (5.8), we find 
$$
\langle p|f\rangle =\frac{1}{\pi i}\int_{-\infty}^{\infty}\frac{\langle
p^{\prime}|(P^+-P^-)|f\rangle }{p-p^{\prime}} dp^{\prime}, \eqno(5.10) 
$$
which is the sum over poles expansion of the Fourier transformed
wavefunction. Eqs. (5.9) and (5.10) can be rewritten in the following way: 
$$
i\langle p|(P^+ -P^- )|f\rangle =\frac{1}{\pi} \int_{-\infty}^{\infty} \frac{%
\langle p^{\prime}|f\rangle }{p-p^{\prime}} dp^{\prime}, \eqno(5.11a)
$$
$$
\langle p|f\rangle =\frac{-1}{\pi} \int_{-\infty}^{\infty} \frac{i\langle
p^{\prime}|(P^+ -P^- )|f\rangle }{p-p^{\prime}} dp^{\prime}. \eqno(5.11b)
$$
\vspace{.15in} Eq. (511a) is a Hilbert transform and Eq. (5.11b) is its
inverse, therefore $i\langle p|(P^+-P^-)|f\rangle $ is the Hilbert transform
of $\langle p|f\rangle $.

\section{SUMMARY AND CONCLUSIONS}

In this work we have defined and discussed the general features of the
incomplete delta functions. They are defind according to 
$$
\delta (x,x^{\prime},P) = \langle x|P|x^{\prime}\rangle , \eqno(6.1) 
$$
where P is a projection operator of quite general nature. They form a very
large class of functions, all of whom satisfy definite relations 
$$
\delta (x,x^{\prime},P) =\int_{-\infty}^{\infty}\delta (x,x^{\prime\prime} 
,P)\delta (x^{\prime\prime} ,x^{\prime},P)dx^{\prime\prime} \eqno(6.2) 
$$
\vspace{.15in} 
$$
\langle x|P|f\rangle =\int_{-\infty}^{\infty}\delta  (x,x^{\prime},P)\langle
x^{\prime}|P|f\rangle dx^{\prime}\eqno(6.3) 
$$
The relations (6.2) and (6.3) are very similar to the ones satisfied by
Dirac's delta functions and form a new type of identities. In section 5 we
dealt with discrete points and by using a similar procedure as in section 2,
we derived the incomplete Kronecker deltas. The definition and an example
are in Eq.(4.10). Similar relations to eqs. (6.2) and (6.3) are the
equations (4.11) and (4.12).

We only touched few examples in section 3, but each example had its own
interesting features. In example 1 the scaling properties were emphasized in
Eq. (3.6). Eq. (3.5) is an interesting identity for waves. In example 3 the
relation (6.3) becomes a Hilbert transform relation between the real and
imaginary parts. A byproduct of the relation (6.2) in example 3 is the
relation (3.15).  The identities obtained can be used to check hypotheses on
functions  that their support is restricted only to a subspace. This
situation exists in signal analysis when the frequencies are restricted to a
bandwidth, or in quantum mechanics when coordinates are limited by an
infinite well potential.

In section 5 dispersion relations for the momenta are obtained and "sums
over poles" are defined and obtained with the aid of differences of
incomplete delta functions. We found this way an expression for the spectrum
of the Hilbert transform.

In this paper we used only the coordinate and momentum bases, but the
formalism is quite general and can accommodate other physical bases as well.
As the number of projection operators and bases is unlimited ${[3],[11]}$ it
seems that there is great potential for new developments in using the
relations (6.2) and (6.3). Presently, for example, we apply them in quantum
mechanical problems in which boundary conditions can be replaced by
orthogonality conditions ${[12]}$.

\section*{ACKNOWLEDGEMENTS}

It is a pleasure to thank Leon Cohen, Arye Friedman, Joel Gersten and Joseph
Malinsky for valuable discussions and encouragement.


\end{document}